\documentclass[review]{elsarticle}

\usepackage[breaklinks]{hyperref}
\usepackage{array}
\usepackage{aeguill}

\usepackage{framed,multirow}
\usepackage{amssymb}
\usepackage{latexsym}
\usepackage{url}
\usepackage{xcolor}
\definecolor{newcolor}{rgb}{.8,.349,.1}

\journal{Advances in Space Research}

\biboptions{authoryear}

\begin{document}

\begin{frontmatter}

\title{Observation and modeling of complex transient structure in heliosphere followed by geomagnetic storm on May 10--11, 2024}

\author{Denis Rodkin \corref{cor1}}
\cortext[cor1]{Corresponding author}
\ead{rodkindg@gmail.com}
\author[1]{Vladislav Lukmanov}
\author[1]{Vladimir Slemzin}
\author[1]{Igor Chashei}

\address{P.N. Lebedev Physical Institute, Leninski prospect 53, Moscow, 119991, Russia}

\begin{abstract}
Complex CME/ICME structures in the solar wind often arising in the heliosphere as a result of interaction between two or more CMEs are very important due to their enhanced geoefficiency, but their modeling is difficult due to lack of observational data outside the solar corona. The outstanding evidence of such complex structure occurred on May 10--11, 2024, when the strongest geomagnetic storm was caused by a series of successive CMEs emerged from the same solar AR~13664. The complex formed from the first four CMEs of the series triggered a drop of the Dst-index to -412 nT. The aim of this study is to consider propagation of these ICMEs in the heliosphere using observations at three stages: at the starting point observed with the LASCO C3 coronagraph, in the middle heliosphere by the IPS method and in situ at the L1 point with the ACE spacecraft. The IPS observations were carried out on May 9 and 10 with the Big Scanning Array radio telescope of the Lebedev Physical Institute (BSA LPI), which enables to build in the scanning mode a 2D map of scintillation index $m^2$ associated with enhancements of the integrated over the line-of-sight (LOS) plasma density over a range of the heliocentric distances 0.4--0.8 AU. Evolution of the ICMEs in the heliosphere was described using a cone model in the self-similar approximation and kinematics according to the Drag-based model. The initial spatial distributions of density at the cone base were determined from the LASCO C3 images of the CMEs at the heights of $\sim~20~R_{sun}$, and then were rescaled according to radial distances to the position corresponding to the time of the BSA LPI observations. It was shown that the modeled distribution of the ICME complex LOS density $N^2$ over heliocentric distance is consistent with distribution of $m^2$. Some discrepancies may characterize variation of the density structure inside the complex due to CME interaction. The modeled mean volume density of the ICME plasma at the L1 position agreed with the in situ ACE measurements, which validates the developed expansion model.
\end{abstract}


\end{frontmatter}

\section{Introduction}
Coronal mass ejections (CMEs) are expanding magnetic ropes or clouds that form as a result of the prominences or filaments eruption in the solar corona. They are visually observed by optical observations and can be accompanied by flares \citep{gosling1990, webb2012}.

The manifestations of CMEs in heliosphere are interplanetary coronal mass ejections (ICMEs). They appear in the corona as a result of plasma eruption with formation and expansion of magnetic ejecta seen by EUV telescopes and then by white light coronagraphs as a CMEs. After reaching the height of $\sim~20~R_{sun}$, they propagate in the heliosphere as ICMEs, large-scale transient structures of magnetized plasma identifiable in the solar wind by significant deviations of the main parameters from the ambient values: proton velocity, density, temperature, and the magnetic field \citep{richardson2004, richardson2010}.

ICMEs often followed by significant disturbances in the Earth's magnetosphere that cause to the emergence of moderate and strong magnetic storms, especially at the solar maximum activity \citep{slemzin2019, temmer2021dec}. To assess the intensity of magnetic disturbances it is often used the low-latitude Dst index (based on magnetic field measurements at four equatorial stations). In this case, storms are called small (-30 nT $\leq$ peak Dst $\leq$ -50 nT), moderate (-50 nT $\leq$ peak Dst $\leq$ -100 nT), and intense (peak Dst $<$ -100 nT) \citep{gonzalez1994}.

Consideration of complex transient structures is of particular interest in the study of geoeffective events. Such structures appear due to interaction of different solar wind flows. Because of interaction, the resulting solar wind magnetic field can strengthen and the duration of its impact on the Earth's magnetosphere increases, which affects the efficiency of storm generation. As a result, complex transients can often lead to the increased geomagnetic activity and the emergence of the most intense magnetic storms \citep{rodkin2020, wang2003, xie2006}.

The prediction of such complex transient structures is complicated by the fact that due to interaction the shape and structure of the participating ICME can be changed. In some cases, it is possible to analyze the interaction of ICMEs using data from several spacecrafts that observe the Sun from different angles \citep{shen2012,soni2023}. Moreover, it can be used 3D magnetohydrodynamic simulations to parametrically evaluate successive interacting ICMEs within a representative heliosphere \citep{koehn2022}. However, such methods are not applicable for regular forecasting, since there is no data on ICME observations in the heliosphere.

One of the methods for regular short-term forecasting of propagating large-scale solar wind disturbances is the interplanetary scintillations (IPS) observation of extragalactic radio sources, which were first discovered by \citet{hewish1964}. The IPS method based on the effect that radiation from radio sources during its passing through the interplanetary medium is distorted due to diffraction on the solar wind density irregularities. This leads to fluctuations in the radiation intensity and emergence of interplanetary scintillations of the radio source. The advantage of the method is the possibility of probing areas of the heliosphere inside the Earth's orbit that are inaccessible to spacecraft. This allows for early detection of solar wind disturbances propagating in the Earth's direction. It is possible to study the global structure of solar wind and estimate its speed using IPS \citep{Chashei2021, tokumaru2024}.

The paper \citep{lukmanov2023} describes the possibility of CMEs detecting before they reach the Earth based on IPS observations at the Pushchino Radio Astronomy Observatory with the Big Scanning Array radio telescope of the Lebedev Physical Institute (BSA LPI). ICMEs can be detected by increase in scintillation at the heliocentric distances from 0.4 to 0.8 AU. In the case of the ICME propagation directed to the Earth, scintillation enhancements are observed in the scanning mode on two-dimensional scintillation index map first at the eastern and then at the western parts around the Sun. The lead-time of the increase in scintillations in relation to the onset of a magnetic storm is about 20 hours on average.

The study considers the complex of a series of ICMEs that caused the intense magnetic storm (with peak Dst $<$ -400 nT) on May 10--11, 2024 \citep{romano2024}. The aim of the study was to identify CMEs as the solar sources of this geoeffective event and to construct a model of their propagation in the heliosphere based on coronagraphic observations near the Sun, IPS observations in the middle heliosphere and in situ measurements in solar wind near the Earth. The results demonstrate validation of the developed model of expansion and possibilities of using the IPS method for studying complex ICME-structures in the heliosphere.

Section 2 describes the data sources and methods used in this study. Section 3 presents the intense magnetic storm of May 10--11, 2024. The ICMEs associated with this storm and their parameters observed in the solar wind with ACE (Advanced Composition Explorer) are presented. The search of CME-sources for ICMEs and calculation of their kinematics is performed. Section 4 discusses the principles of ICME observations using the IPS method on the BSA LPI and its observation data for May 9 and 10, 2024. Section 5 presents the calculation of the spatial distribution of CMEs LOS density based on LASCO data, its recalculation to the time of observations by the BSA LPI, and a comparison with the spatial distribution of the scintillation index. Moreover, it is presented a calculation of the plasma volume density for the first ICME near the Earth (at point L1) in comparison with the measurement data from ACE. The Conclusion summarizes the results of this study.
\section{Data and methods}
We used solar wind parameters from ACE spacecraft (L1 Lagrange point) to identify ICMEs associated with geomagnetic storm on May 10--11, 2024\footnote{https://izw1.caltech.edu/ACE/ASC/browse/view\_browse\_data.html}. ICMEs  boundaries were estimated by intervals of increased proton speed and density, increased magnetic field strength, a drop in proton temperature (below expected for normally-expanding solar wind with the same speed), and a low proton plasma beta value (the ratio of the plasma pressure to the magnetic pressure) \citep{zurbuchen2006, richardson2010}. The comparison of ACE data with the drop in the geomagnetic Dst-index (from the World Data Center for Geomagnetism\footnote{https://wdc.kugi.kyoto-u.ac.jp/}) showed that the strong geomagnetic disturbance on May 11, 2024 was caused by ICMEs recorded near the Earth on May 10--11, 2024.

The first step was to determine the time intervals at the Sun when one or several CMEs could produce the given ICMEs. We calculated these intervals using the average ICMEs speeds in the ballistic approximation \citep{nolte1973} with an accuracy of $\pm 24~h$. The second step was to select CMEs directed at Earth. For this purpose, we used the data from CDAW\footnote{https://cdaw.gsfc.nasa.gov/CME\_list/} and CACTus\footnote{https://www.sidc.be/cactus/catalog.php} catalogs. 7 Halo-CMEs that occurred in the selected time interval and were directed toward Earth has been established as the most probable sources of our ICMEs.

To describe propagation of CMEs/ICMEs in interplanetary space, the Drag-Based Model (DBM, \citep{cargill2004, vrsnak2013, vrsnak2021}) was used due to its simplicity and low computational cost. This model assumes, that from a certain distance from the Sun (typically, at $R \geq 20~R_{sun}$), where the Lorentz and gravity forces become negligible, the ICME dynamics is governed by magnetohydrodynamic drag produced by interaction of the ICME plasma with the interplanetary ambient (background)  solar wind. The basic version of DBM founded on the equation for the drag force acceleration:
\begin{eqnarray}
\frac{dv}{dt} = -\gamma(v-w)|v-w|, v(t)= \frac{v_0-w}{1\pm\gamma(v_0-w)t}+w,
\end{eqnarray}
where $v$ is the ICME speed along the designated direction, $w$ represents the ambient solar wind speed in front of ICME, $v_0$ is the CME speed at the distance $20~R_{Sun}$ and $\gamma$ is the drag parameter. The typical error of the DBM method in its basic form can be estimated as $\pm 10~h$ (in time) and $\pm 50~km~s^{-1}$ (in speed) \citep{vourlidas2019, dumbovic2021}. These errors depend on uncertainties in the measured initial CME speed, the ambient solar wind speed along the path of ICME and unknown drag coefficient. Data from CDAW and CACTus catalogs were used to determine the CMEs speeds. Background speeds were estimated using ACE data.

Interplanetary scintillation data were obtained at the BSA LPI radio telescope \citep{shishov2016}. Observations on the upgraded BSA LPI have been performed since 2014 at a central frequency of 111 MHz in the 2.5 MHz frequency band. The Fresnel $a_f$ scale for this frequency is on the order of several hundred kilometers.

During observation of scintillation at the BSA LPI in monitoring mode about 5000 scintillating radio sources pass through the radiation pattern of the radio telescope per day. To predict space weather, observations are used in the daytime (usually from 7 to 16 o'clock), when, due to rotation of the Earth, the area around the Sun is scanned, corresponding to a heliocentric distance of the order of 1 AU, from which the most suitable part with a minimum of interference at heliocentric distances from 0.4 to 0.8 AU is selected for analysis. The entire observable area of the sky is divided into $3^{\circ}~\times~3^{\circ}$ pixels, each of which contains up to $\sim 10$ observable radio sources. For all pixels, the scintillation index averaged over all the sources in the pixel is calculated using the formula:
\begin{eqnarray}
m^2 = \frac{\langle (I(t)-\langle I \rangle)^2 \rangle}{\langle I \rangle ^2},
\end{eqnarray}
where $I(t)$ is the measured radiation flux density of the radio source as a function of time, $\langle I \rangle$ is its time average value. The value of $m^2$ is proportional to the square of plasma density fluctuations on the Fresnel scale $\langle \delta N_{af}^2 \rangle$, which is integral over the line of sight to this source. It is usually assumed that the value $\langle \delta N_{af}^2 \rangle$ is proportional to the mean square of the integral plasma density at the proximate point of the line-of-sight to the radio source $\langle N^2 \rangle$ \citep{lukmanov2022dec}, therefore, large-scale amplifications in the spatial distributions of the scintillation indices will correspond to disturbances of the solar wind with increased concentration. Based on observational data, two-dimensional maps of the distribution of the scintillation indices squared are produced daily.

To interpret the scintillation maps obtained by the BSA LPI and related to appearance of the ICME complex in the field of view we used the CME density maps calculated from the images registered by the LASCO C3 coronagraph at the SOHO observatory \citep{brueckner1995}. Expansion of the ICMEs in the heliosphere was described by the cone model in the self-similar approximation using kinematics provided with DBM. The ICME density maps corresponding to the IPS maps were obtained by rescaling the initial CME density maps at the heights about $20~R_{Sun}$ to the heights where the ICMEs appeared in the BSA LPI field of view during the observation session.
\section{CME-sources and kinematics of the complex transient structure}
The intense G5 (Kp=9) magnetic storm began on May 10, 2024 (SWPC NOAA\footnote{swpc.noaa.gov}). The Dst-index value at the maximum of the storm dropped to -412 nT, and the recovery time was more than 3 days.

\begin{figure}
\centering{\includegraphics[width=0.8\linewidth]{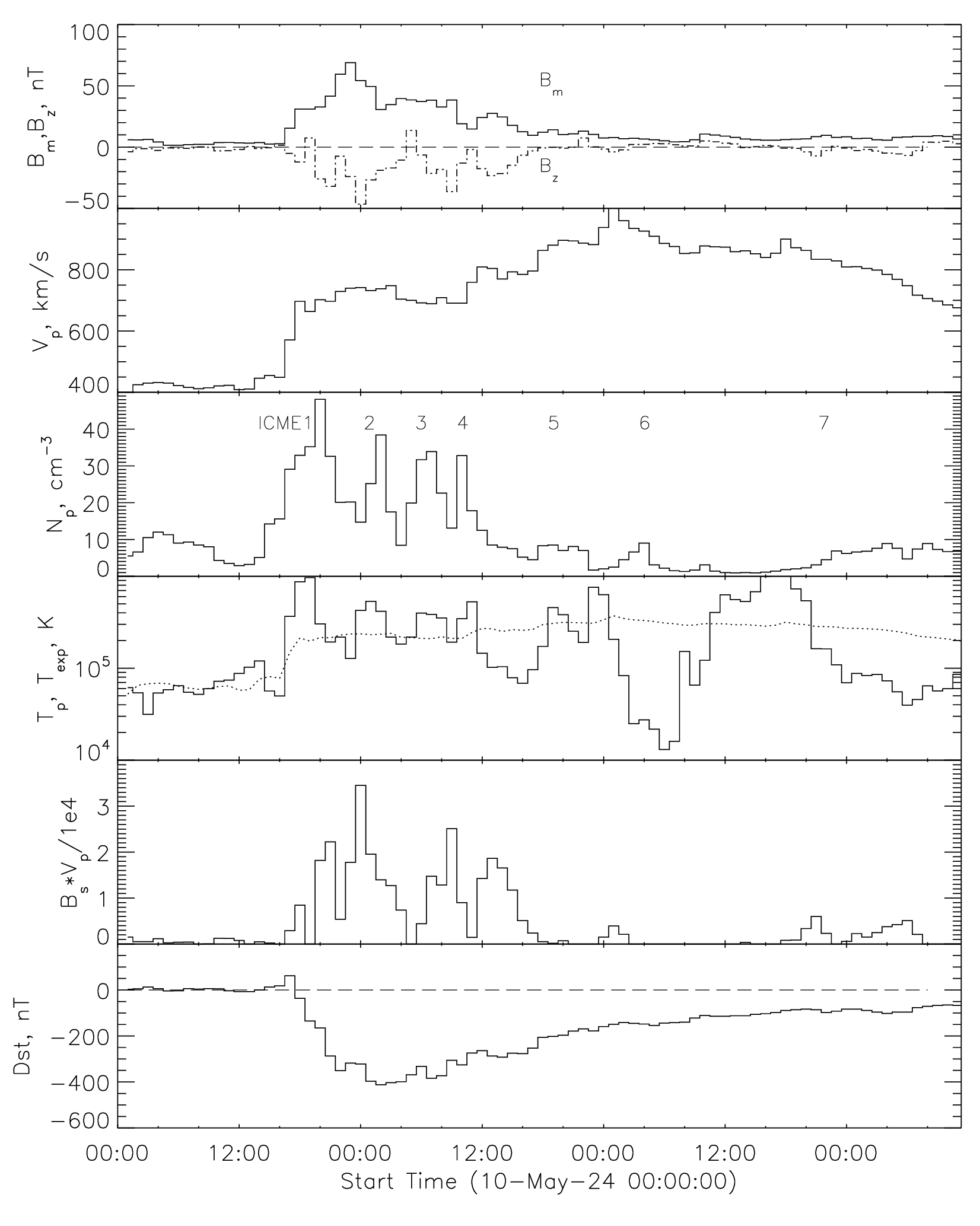}
\caption{Parameters of the solar wind related to the geomagnetic storm on May 10--11, 2024: from top to bottom -- the amplitude $B_m$ and the component $B_z$ of the magnetic field; proton speed ($V_p$); density ($N_p$) and temperature ($T_p$) (the dotted line shows the expected temperature -- $T_{exp}$); dawn-dusk electric field ($B_s \cdot V_p$) (the values are divided by $10^4$) value of Dst-index. The numbers on the third graph indicate the ICME number.}
\label{figure:1}}
\end{figure}

There were identified 7 ICMEs associated with this magnetic storm (Fig.~\ref{figure:1}) using SW parameters obtained by ACE spacecraft. These ICMEs can be combined into a single complex structure based on their impact on the Earth. A detailed investigation of such structures can be found, for example, in \citep{lugaz2017, rodkin2018}.

The first 4 ICMEs were associated with the strongest drop in Dst-index during the period of May 10--11 (see Fig.~\ref{figure:1}). The 3 remaining ICMEs were associated with a long post-storm recovery phase. It should be noted that a strong dip in Dst-index corresponds to the smallest value of the Bz magnetic field component (or the largest value of the southern component $B_s = |B_z < 0|$) and the largest value of dawn-dusk electric field ($v \cdot B_s$). Changes in these parameters are closely related with the geoeffectiveness of the solar wind \citep{zhang2007, shen2017, rodkin2020}.

In this study, we focused on the first 4 ICMEs. For them we estimated the time interval of their CME-sources in the period of May 8--9, 2024 using ballistic approximation. There were 4 successive Halo-CMEs in this time period (according to LASCO data), which were accompanied by powerful M and X class flares in the active region AR~13664 (according to SolarMonitor data). The evolution of AR~13644 during its passage across the solar disk before the onset of the flares and CMEs responsible for the geomagnetic storm described in \citep{romano2024}. Various features of the X-class flares, which occurred in the AR~13664 during May 8--15, 2024 presented in \citep{li2024}.

Further, we used DBM to obtain the more accurate near-Earth values of arrival time and velocity for 4
selected CMEs. Table~\ref{table:1} presents the input parameters used in this model. The ambient speed value and drag parameter were selected to obtain a minimum discrepancy between the calculated (DBM) and in-situ (ACE data) ICMEs arrival time. The ambient speed ($w$) was taken in the range from 450 to 700 $km~s^{-1}$ (see the graph of proton speed in Fig.~\ref{figure:1}). The drag parameter ($\gamma$) was chosen to be from 0.2 to $0.4 \cdot 10^{-7} km^{-1}$, as the most typical \citep{calogovic2021, rodkin2023, rodkin2024}.

The CME speed in DBM was calculated using values obtained from the CACTus and CDAW catalogs, assuming 
that the longitudinal average CME speed toward the Earth is equal to its average expansion speed ($V_{rad} = V_{exp}$). This assumption was based on the research results conducted in \citep{gopalswamy2009jan, michalek2009, makela2016}, where the authors proposed the equation for «full ice-cream cone model»: 
\begin{eqnarray}
V_{rad} = \frac{1}{2}(1+\cot \beta)V_{exp},
\end{eqnarray}
where $\beta$ is a half of the cone opening angle. According to the data from \citep{makela2016}, the average Halo-CME width (for 2011--2012) was equal to $\approx 86.6^{\circ}$ with a standard error of $\approx 6.6^{\circ}$. So, we obtained $V_{rad}/V_{exp} = 1.03 \pm 0.06$. Thus, it can be considered that $V_{rad} \approx V_{exp}$ within $10\%$ error.

\begin{table}
\centering
\caption{Values of CME parameters used to obtain their near-Earth arrival time (in UT) and velocity from DBM. $T_0$ is the time of CME appearance, $T_{r20}$ is the time of CME arrival at $20~R_{Sun}$, $V_{r20}$ is the velocity of CME at $20~R_{Sun}$, $\gamma$ is the drag parameter, $w$ is the background velocity, $T_{Earth}$ is the estimated time of CME arrival to Earth according to DBM, $V_{Earth}$ is the estimated velocity of CME at Earth according to DBM. Values in bold are based on CDAW catalog data. Values in italics are based on CACTus data}
\label{table:1}
\begin{tabular}{|l|l|l|l|l|l|l|l|}
\hline
CME & $T_0$ & $T_{r20}$ & $V_{r20},$ & $\gamma,$ & $w,$ & $T_{Earth}$ & $V_{Earth}$ \\
 & & & $km~s^{-1}$ & $10^{-7} km^{-1}$ & $km~s^{-1}$ & & $km~s^{-1}$ \\
\hline
1 & \textit{May 8, 2024} & \textit{May 8, 2024} & \textit{856} & 0.2 & 450 & May 10, 2024 & 606 \\
 & \textit{4:00} & \textit{8:17} & & & & 15:18 & \\
\hline
2 & \textit{May 8, 2024 } & \textit{May 8, 2024} & \textit{750} & 0.3 & 600 & May 10, 2024 & 680 \\
 & \textit{12:48} & \textit{17:42} & & & & 23:29 & \\
\hline
3 & \textbf{May 8, 2024} & \textbf{May 9, 2024} & \textbf{954} & 0.3 & 550 & May 11, 2024 & 677 \\
 & \textbf{22:24} & \textbf{2:00} & & & & 3:47 & \\
\hline
4 & \textbf{May 9, 2024} & \textbf{May 9, 2024} & \textbf{1224} & 0.3 & 650 & May 11, 2024 & 813 \\
 & \textbf{9:24} & \textbf{11:45} & & & & 4:24 & \\
\hline
\end{tabular}
\end{table}

Table~\ref{table:2} presents a comparison between the calculated (DBM) and measured values (ACE data). The average deviation for 4 CMEs in velocity was $\approx 36~km \cdot s^{-1}$, and in time was $\approx 5.2~h$.

\begin{table}
\centering
\caption{Comparison between calculated (according to DBM) and measured (according to ACE data) values of ICMEs arrival times and speeds}
\label{table:2}
\begin{tabular}{|l|l|l|l|l|l|l|}
\hline
CME & $T_{DBM}$ & $T_{ACE}$ & $\Delta T, h$ & $V_{DBM},$ & $V_{ACE},$ & $\Delta V,$ \\
 & & & & $km~s^{-1}$ & $km~s^{-1}$ & $km~s^{-1}$ \\
\hline
1 & May 10, 2024 & May 10, 2024 & 4.9 & 608 & 688 & 80 \\
 & 13:56 & 18:50 & & & & \\
\hline
2 & May 10, 2024 & May 11, 2024 & 4.07 & 681 & 724 & 43 \\
 & 22:16 & 2:20 & & & & \\
\hline
3 & May 11, 2024 & May 11, 2024 & 3.43 & 680 & 688 & 8 \\
 & 2:34 & 6:00 & & & & \\
\hline
4 & May 11, 2024 & May 11, 2024 & 8.57 & 816 & 802 & 14 \\
 & 3:22 & 12:00 & & & & \\
\hline
\end{tabular}
\end{table}

\begin{figure}
\centering{\includegraphics[width=0.7\linewidth]{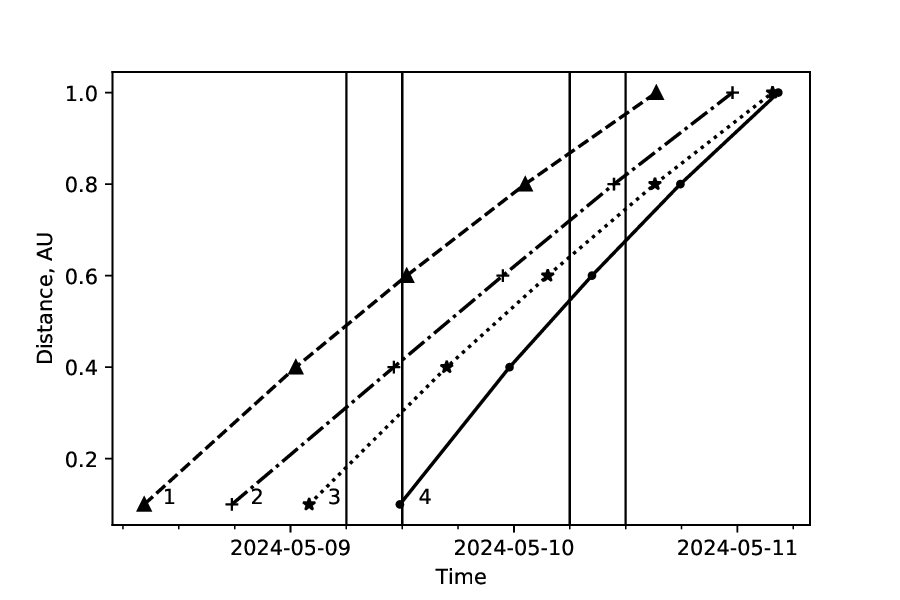}
\caption{The movement of the first 4 CMEs up to 1.0 AU based on the results of DBM calculations. The vertical solid lines indicate the observational time intervals at the BSA LPI.}
\label{figure:2}}
\end{figure}

Figure~\ref{figure:2} presents the predicted motion of 4 Halo-CMEs in the heliosphere on the Sun-Earth line, obtained from the DBM calculation. It is clear that they follow each other compactly and arrive at the Earth at intervals of approximately 8, 4 and 1 hour. So close proximity could enhance the overall geoefficiency of such structure. For example, in \citep{koehn2022} the authors have shown how the interaction of two moderate CMEs between the Sun and the Earth can translate into extreme conditions at the Earth.

The equality condition of the average longitudinal and expansion velocities corresponds to the self-similar ICME size increase and means that the average transverse transient size is approximately equal to the distance to the Sun. It follows from the graph in Fig.~\ref{figure:2} that during the observation period on May 9 from 06:00 to 12:00 UT, ICMEs 1--3 were in the BSA LPI field of view, but ICMEs 2 and 3 were too close to the Sun, so they had transverse size less than 0.4 AU and could not be identified by BSA LPI. During the observation period on May 10, all 4 ICMEs passed through the BSA LPI field of view.
\section{IPS observations of CMEs at BSA}
\begin{figure}[h]
\begin{minipage}[h]{0.49\linewidth}
\center{\includegraphics[width=0.9\linewidth]{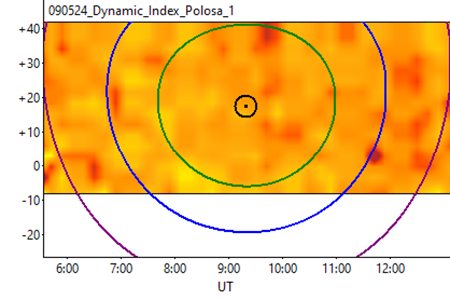} \\ a)}
\end{minipage}
\hfill
\begin{minipage}[h]{0.49\linewidth}
\center{\includegraphics[width=0.9\linewidth]{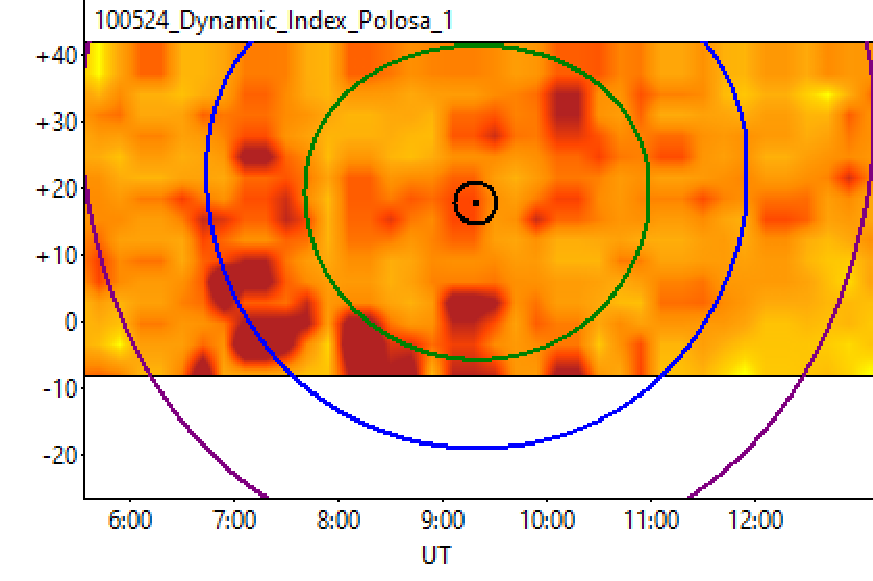} \\ b)}
\end{minipage}
\caption{Observational dynamic maps of scintillation indices: a) for May 9, 2024; b) for May 10, 2024.}
\label{figure:3}
\end{figure}

Figure~\ref{figure:3} shows the dynamic scintillation index maps for May 9 and 10, 2024, calculated by dividing the scintillation index map for a given day by the map for the previous day pixel by pixel. This pixel-by-pixel division is an analogue of the g-index, in which the observed value of the scintillation index is divided by the value of the scintillation index expected for a given source at a given elongation. Use of the g-index is generally accepted when searching for disturbances in turbulent solar plasma. However, as shown in \citep{lukmanov2022feb}, the dependence of the scintillation index on elongation can vary from year to year and therefore the g-index is a convenient parameter for scientific research of turbulent plasma in general. At the same time, an increase in the scintillation index in a pixel when comparing the current day with the previous day shows a possible disturbance of the solar wind associated with appearance of a high-density plasma cloud on the line of sight. If this increase in the scintillation index is observed simultaneously in many pixels, the probability of detecting a disturbance becomes very high.

The Universal time (UT) is marked on the horizontal axis on the maps, declination is marked on the vertical axis. The orange color on the maps corresponds to steady conditions (the scintillation index remained at about the same level), red and shades of red -- enhancement of scintillation compared to the previous day, yellow -- weakening of scintillation. The arcs heliocentric distances of the modulating layer (the proximate point of the line of sight to the radio source) are plotted: green -- 0.4 AU, blue -- 0.6 AU, purple -- 0.8 AU. It can be seen that the intensification of scintillation began on the day before the start of the magnetic storm on May 11 in the area of heliocentric distances of 0.4 -- 0.8 AU.

In the data for May 9, in the range of heliocentric distances of 0.4 -- 0.6 AU, a weak signal is observed at the background level, which is apparently associated with the passage of the outer part of the cloud of ICME 1, and it was not quantitatively analyzed due to the small signal-to-noise ratio of less than 1. The map for May 10 reflects the position of the plasma cloud at different points of measurement. Table~\ref{table:3} provides information on the average values of the scintillation indices for May 9--10 in different intervals of heliocentric distances, as well as the corresponding values in the absence of solar wind disturbances. It should be noted that the background level was determined by the average signal level on the quiet days (May 6 and 7), when there were no disturbances associated with the solar wind.

Note that IPS amplifications on May 10, 2024 were also recorded in the ISEE data of Nagoya University at a frequency of 327 MHz \citep{hayakawa2024}.

\begin{table}
\centering
\caption{Average indices of scintillation for May 9 and 10 and background level in different intervals of heliocentric distances}
\label{table:3}
\begin{tabular}{|l|l|l|l|l|}
\hline
Ring radii, AU & Signal from ICME & Signal from ICME & Background & Background \\
 & on May 9 & on May 10 & signal & r.m.s. variance \\
\hline
0.40 -- 0.45 & 0.332 & 0.434 & 0.301 & 0.018 \\
\hline
0.45 -- 0.50 & 0.370 & 0.527 & 0.423 & 0.063 \\
\hline
0.50 -- 0.55 & 0.455 & 0.545 & 0.416 & 0.044 \\
\hline
0.55 -- 0.60 & 0.466 & 0.528 & 0.514 & 0.042 \\
\hline
0.60 -- 0.65 & 0.512 & 0.577 & 0.519 & 0.026 \\
\hline
0.65 -- 0.70 & 0.490 & 0.554 & 0.441 & 0.023 \\
\hline
0.70 -- 0.75 & 0.456 & 0.493 & 0.413 & 0.038 \\
\hline
0.75 -- 0.80 & 0.449 & 0.427 & 0.414 & 0.040 \\
\hline
\end{tabular}
\end{table}
\section{ICME density in the heliosphere in comparison with the BSA LPI data}
The maps of the ICME spatial density distributions integrated over the line-of-sight (LOS) analogous to the BSA LPI maps of scintillation index were created based on the CME data from the CDAW catalogue and the data from calculations of kinematics in the heliosphere. The calculations were carried out under the following assumptions.
\begin{enumerate}
  \item The ICMEs were modeled as cones with density distribution in their base taken from the most contrast images in the LASCO C3 coronagraph at heights about $20~R_{Sun}~(H_{20})$. 
  \item The expansion of the ICMEs was assumed to be self-similar with preservation of shape in the base. The linear dimensions R of all ICME fragments in the picture plane were enlarged with respect to the initial ones $R_{20}$ at $H_{20}$ proportionally to the height from the solar surface $H$: $R = R_{20}~\cdot~(H/H_{20})$. 
  \item The ICMEs constituting the complex structure propagated in the heliosphere towards the Earth radially according to the Drag-based model. Influence of their interaction was taken into account by the choice of the model parameters corresponding to the in situ solar wind measurements. 
\end{enumerate}

According to the studies in the papers \citet{gopalswamy2009mar, nieves2013, zhang2022, cecere2023}, in some cases CMEs at their paths in the corona below the height of $\sim~11~R_{Sun}$ decline from the initial radial directions under influence of the neighboring large coronal holes or current sheets, especially, in the solar minima. In our case these effects should be small because the studied ICMEs propagated in the period of the solar maximum at the heights above $20~R_{Sun}$, where external magnetic fields become weaker and the movement of ICMEs becomes rectilinear with a fixed angular distribution of transverse velocities. Thus, application of the self-similar approximation of ICMEs expansion in the heliosphere is quite justified, that is why it is widely used in the ICME modeling including the MHD calculations \citep{scolini2018, temmer2021jan, dai2022}.

\begin{figure}[h]
\begin{minipage}[h]{0.49\linewidth}
\center{\includegraphics[width=0.9\linewidth]{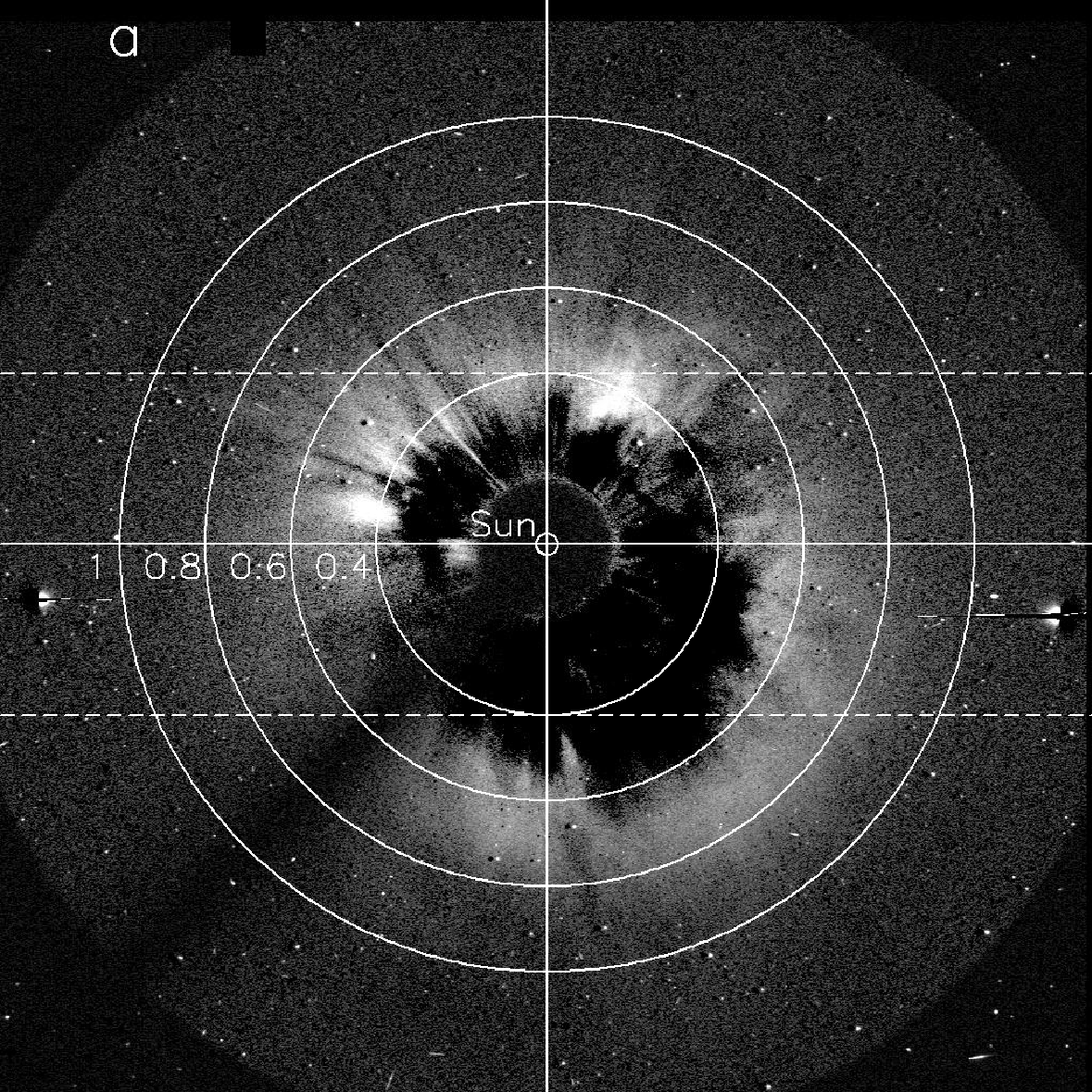} \\ }
\end{minipage}
\hfill
\begin{minipage}[h]{0.49\linewidth}
\center{\includegraphics[width=0.9\linewidth]{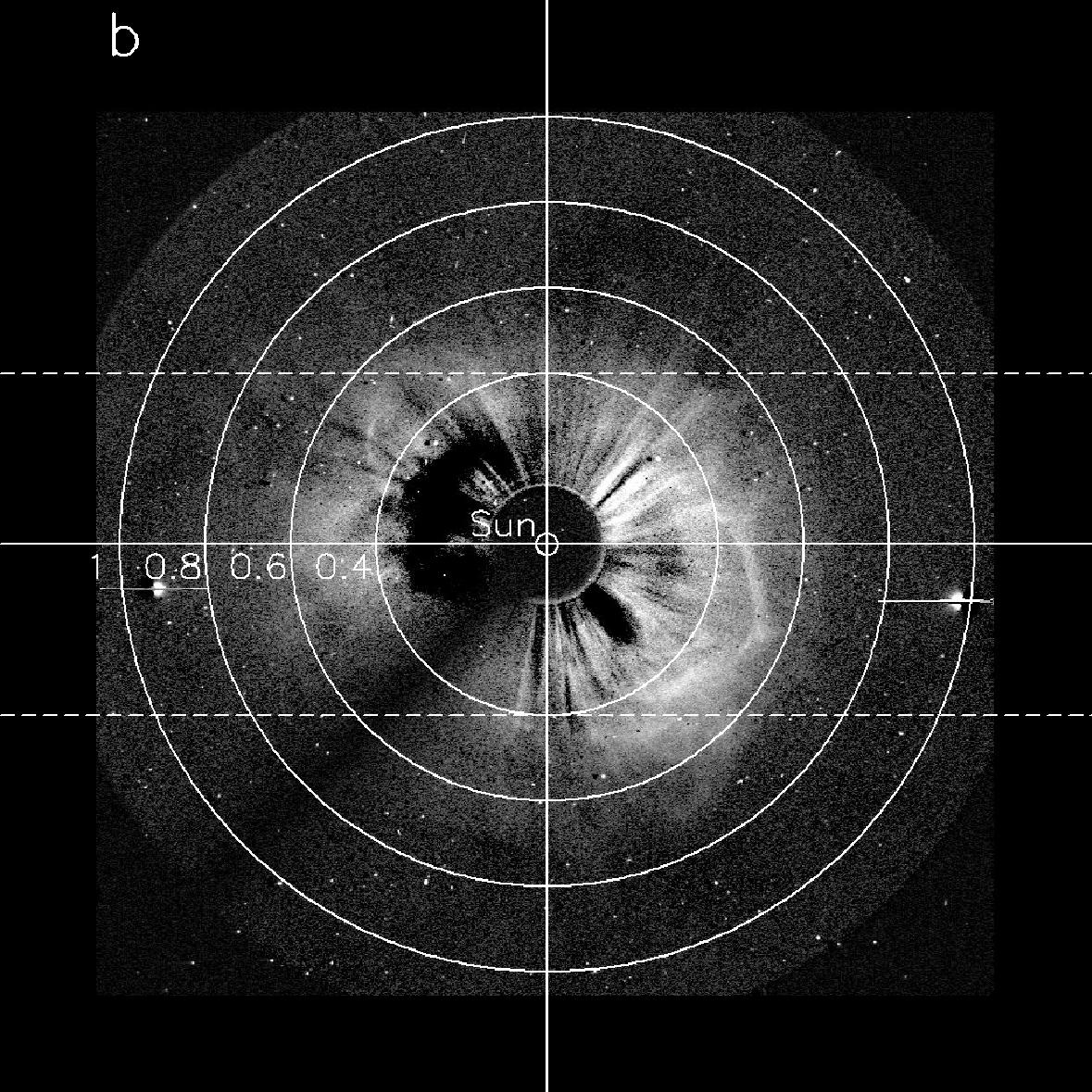} \\ }
\end{minipage}
\caption{a) The image of CME 1 obtained by the LASCO C3 coronagraph on May 8, 2024 at 11:42 UT scaled to the BSA LPI field of view at the observational time on May 10, 2024 09:00 UT; b) the image of CME 3 obtained on May 8, 2024 at 22:30 UT scaled to the BSA LPI field of view at the same observational time. The horizontal dashed lines delimit the central areas of the rings, on which the densities were averaged.}
\label{figure:4}
\end{figure}

In frames of the developed model the ICME images were scaled in proportion with increase of the height above the Sun from the initial ($H_{20}$) value corresponding to the LASCO C3 measurement up to the value corresponding to observation of scintillations by BSA LPI ($H_{BSA}$). Taking into account difference in the picture angular scales, the expansion coefficient is:
\begin{eqnarray}
K_{exp} = \frac{H_{BSA}}{H_{20}}~\cdot~\frac{R_{Sun~BSA}}{R_{Sun~LASCO}},
\end{eqnarray}
where $R_{Sun~BSA}$ and $R_{Sun~LASCO}$ -- the solar radius (in pixels) in the BSA LPI field of 
view and coronagraph, correspondingly. Fig.~\ref{figure:4} shows images of ICME 1 (a) and ICME 3 (b) scaled to the BSA LPI field of view.

\begin{figure}[h]
\begin{minipage}[h]{0.49\linewidth}
\center{\includegraphics[width=0.9\linewidth]{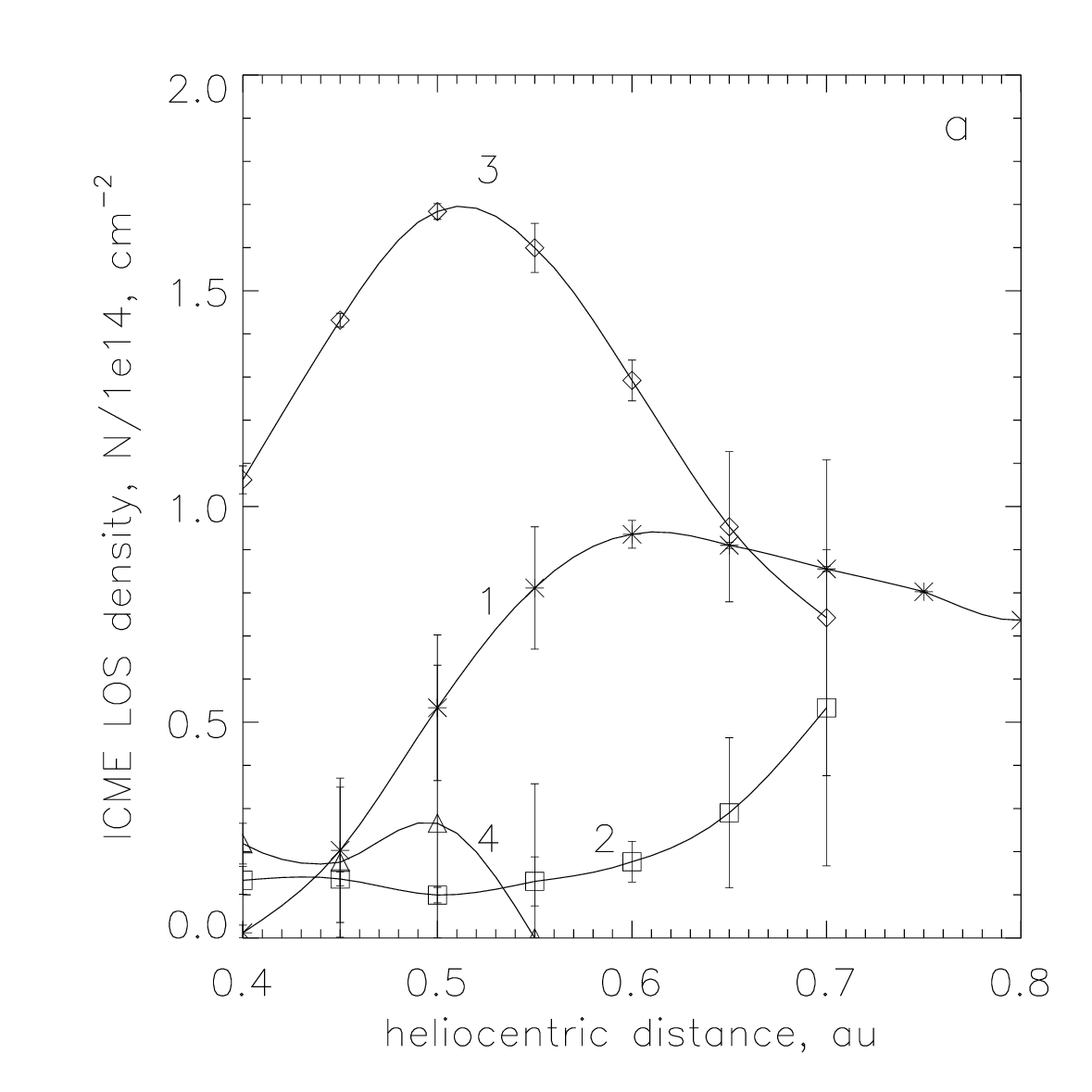} \\ }
\end{minipage}
\hfill
\begin{minipage}[h]{0.49\linewidth}
\center{\includegraphics[width=0.9\linewidth]{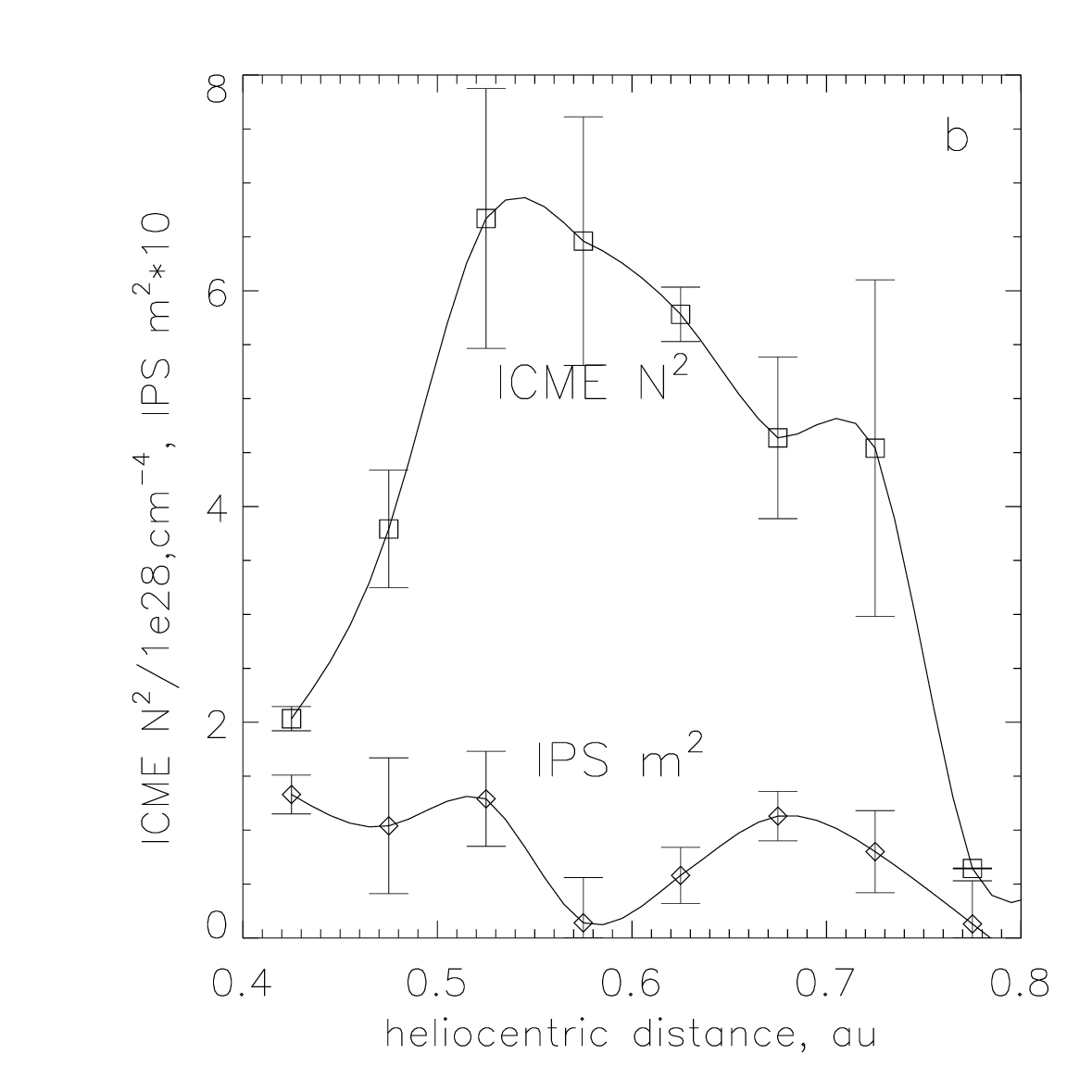} \\ }
\end{minipage}
\caption{The modeled spatial distributions of the integral (by LOS) plasma density for ICMEs 1--4 in the BSA LPI field of view ($n$) divided by $10^{14}~cm^{-2}$ (left panel) and comparison of the distribution of the summed ICMEs density in a square ($n^2$) with the distribution of the scintillation index from the BSA LPI data (right panel).}
\label{figure:5}
\end{figure}

After establishing a correspondence between BSA LPI and coronagraph field of view, the intensities in the LASCO C3 CME image were selected by the annular segments rescaled back from the BSA LPI field of view with the step of 0.05 AU accounting for the field of view limitations above and below shown in Fig.~\ref{figure:4} by the dashed lines. Then, the masses and mean densities of the CME plasma in each segment integrated over the LOS were calculated using the standard SolarSoft procedures. Because BSA LPI forms the maps of scintillation index in the scanning mode during the interval from 06:00 to 12:00 UT each day, it leads to temporal shift of the IPS picture. In order to take into account this shift, calculations of CME densities from the LASCO C3 images were fulfilled in three cases associated with three BSA LPI observation times 06:00, 09:00 and 12:00 UT. Then the results were averaged, and difference between maximum and minimum values in each segment was treated as the systematic error.

At the next step the densities in the BSA LPI segments were defined from those of conjugated LASCO C3 values. According to the work \citet{temmer2022}, in which they analyzed a series of 40 ICME events observed on the Helios spacecraft in 1975--1981 and 5 events registered on the Parker solar probe in 2020--2021, variation of the ICME density (“magnetic ejecta”) with distance to the Sun in the range 0.03--1.6 au can be approximated by the dependence $N_1/N_0 = (L_1/L_0)^{-2.4}$, and the ICME LOS size as $S_1/S_0 = (L_1/L_0)^{0.78}$. As a result, the ICME density integrated over LOS varies with distance as:
\begin{eqnarray}
\frac{N_1~\cdot~S_1}{N_0~\cdot~S_0}~=~(\frac{L_1}{L_0})^{-1.62},
\end{eqnarray}
where position 0 corresponds to the LASCO CME image, position 1 -- to the BSA LPI map. To take into account compression of the ambient plasma in front of the ICME leading edge ("sheath"), which at the distances larger than 0.09 AU can exceed the main ICME density, we introduced the additional coefficient: 
\begin{eqnarray}
K_{sht} = 1 + 0.46~\cdot~(\frac{L_{BSA}}{L_{AU}})^{2.8},
\end{eqnarray}
which equals to relation of the product of density and thickness of a sheath to that of the ICME body as a function of distance in the units of 1 AU.

In Figure~\ref{figure:5} (left panel) shows spatial distributions of integrated densities for ICMEs 1--4 in the heliosphere calculated for the distances corresponding to the BSA LPI observations on May 10, 2024. The right panel demonstrates a comparison of the modeled spatial distribution of the total LOS density in a square $N^2$ summed over ICMEs 1--4 with the measured distribution of scintillation index. In general, there is a similarity between these dependencies with some differences in the details: discrepancies in the positions and magnitudes of maxima and minima in both curves. These differences may be due to changes of the ICMEs densities as a result of their interaction or rotation of the magnetic flux rope \citep{liu2018}.

We validated the ICME expansion model by comparison of the calculated ICME 1 mean volume density in the point L1 with the measurements in situ on ACE. To do that, we used the value of the mean LOS density at the height 195 solar radii, which corresponds to the BSA LPI observations, $N_{LOS}(BSA) = 6~\cdot~10^{13}~cm^{-2}$. According to the work \citep{temmer2022}, the LOS density was scaled to the L1 point (at the height of $211~R_{Sun}$) with the coefficient $(211/195)^{-1.62} = 0.87$. Then, by adding the sheath, the density should increase in 1.34 times giving in total the value $N_{LOS}(L1) = 7~\cdot~10^{13}~cm^{-2}$. From the ACE data, the length of ICME 1 with the sheath is $0.12 AU = 1.85~\cdot~10^{12}~cm$. Thus, it follows, that the mean volume density of the ICME 1 plasma in the L1 point equals to $N_e = 38~cm^{-3}$ (for electrons), which corresponds to the proton density $N_p = 0.9~\cdot~N_e = 34~cm^{-3}$. From the ACE data (Fig.~\ref{figure:1}), the maximum value of the proton volume density in the peak of ICME 1 equals to $48~cm^{-3}$, the mean density -- $24~cm^{-3}$. Thus, for the event under consideration, the density expansion model gives a quantitative result that is consistent with the measurements, which justifies the underlying assumptions. Similar results were obtained in the work \citet{temmer2021jan} by analysis of a series of ICMEs 2010--2011 using the self-similar model GCS (Graduated Cylindrical Shell).
\section{Conclusion}
The work carried out a study of propagation in the heliosphere a series of ICMEs followed by a strong geomagnetic storm on May 10--11, 2024. According to the ACE solar wind data and kinematical consideration, it was determined that the storm was caused by a complex of successive ICMEs, which were the heliospheric manifestation of 4 Halo-CMEs emerged from the same solar active region (AR~13664) on May 8--9. The ICMEs constituting this complex were observed by different instruments in three positions: 1) at the heights of $\sim~20~R_{sun}$ they were seen as CMEs with the LASCO C3 coronagraph, 2) in the middle heliosphere the complex was observed with BSA LPI by the IPS method, 3) in situ at the L1 point the ICMEs were detected on the ACE spacecraft. To interpret these observations, we developed the model, which enables to calculate a spatial 2D-distribution of the LOS density at different heights from the Sun. The ICME density distributions in the heliosphere were calculated for the heights corresponding to the time of BSA LPI observations on May 10, 2024 using a cone model in the self-similar expansion approximation with a base in the form of a density distribution followed from the LASCO C3 images. Comparison of the modeled spatial 2D-distribution of the summed LOS density of all four ICMEs in a square with the BSA LPI scintillation index averaged over the same annular segments shows their similarity, which conforms to the theory of the IPS method. The mean volume density of the ICME plasma scaled to the L1 position agreed with in situ measurements, which validates the developed expansion model. These results demonstrate that IPS method of studying ICMEs in the heliosphere is effective not only for single transients by analogy with optical coronagraph in the corona, but also for investigation of complex interactive structures.
\section*{Acknowledgments}
The authors are grateful to Tyulbashev S.A. for useful discussions and provision of IPS data. We express our gratitude to the BSA LPI technical group for providing observations of interplanetary scintillations. This paper also uses data from the CACTus (by SIDC at the Royal Observatory of Belgium) and CDAW (by NASA and The Catholic University of America in cooperation with the Naval Research Laboratory) CME catalogs. The authors thank the ACE and SOHO research teams for their open data policy. SOHO is a project of international cooperation between ESA and NASA. We also thank the SolarMonitor team\footnote{https://www.solarmonitor.org/} for near-realtime and archived information on active regions and solar flares.
\bibliography{refs}

\end{document}